# DISTRIBUTION-FREE CUMULATIVE SUM CONTROL CHARTS USING BOOTSTRAP-BASED CONTROL LIMITS

By Snigdhansu Chatterjee[1] and Peihua Qiu[2]

*University of Minnesota and University of Minnesota*

This paper deals with phase II, univariate, statistical process control when a set of in-control data is available, and when both the in-control and out-of-control distributions of the process are unknown. Existing process control techniques typically require substantial knowledge about the in-control and out-of-control distributions of the process, which is often difficult to obtain in practice. We propose (a) using a sequence of control limits for the cumulative sum (CUSUM) control charts, where the control limits are determined by the conditional distribution of the CUSUM statistic given the last time it was zero, and (b) estimating the control limits by bootstrap. Traditionally, the CUSUM control chart uses a single control limit, which is obtained under the assumption that the in-control and out-of-control distributions of the process are Normal. When the normality assumption is not valid, which is often true in applications, the actual in-control average run length, defined to be the expected time duration before the control chart signals a process change, is quite different from the nominal in-control average run length. This limitation is mostly eliminated in the proposed procedure, which is distribution-free and robust against different choices of the in-control and out-of-control distributions.

**1. Introduction.** The problem of univariate, phase II statistical process control (SPC) may be described as follows: A sequence of independent random variables $\{X_n, n \geq 1\}$ on the real line is observed, such that $X_1, \ldots, X_{t_0}$ follow a given distribution $F$ (called an "in-control" distribution) and $X_{t_0+1}$, $X_{t_0+2}, \ldots$ follow another distribution $G$ (called an "out-of-control" distribution), where $F \neq G$. The major objective of SPC techniques is to detect such a distributional shift as soon as possible.

Received June 2007; revised May 2008.

[1]Supported in part by a grant-in-aid from the University of Minnesota.

[2]Supported in part by an NSF grant.

The theory and methods of SPC have traditionally developed from industrial statistics roots, such as quality specifications. In modern times, while quality enhancement still remains a major field of applications, SPC has found many other applications. For instance, SPC is widely used in health care monitoring [Steiner, Cook and Farewell (1999)], detection of genetic mutation [Krawczak et al. (1999)], credit card and financial fraud detection [Bolton and Hand (2002)], and insider trading in stock markets [Meulbroek (1992)]. In such applications, process distributions are often multimodal, skewed, or heavy tailed.

In traditional SPC, $F$ is usually assumed to be a known Normal distribution, and $G$ is a different Normal distribution. When a shift in the mean of $F$ is the major concern, the minimax sequential probability ratio test known as the "Cumulative Sum Control Chart" (CUSUM chart hereafter) is the dominant technique for detecting such a shift [cf. Page (1954)]. To detect an upward shift, the CUSUM $C_n$ is defined by $C_0 = 0$, and

$$(1) \qquad C_n = \max\left(C_{n-1} + X_n - k, 0\right) \qquad \text{for } n \geq 1,$$

where $k \geq 0$ is a pre-specified *allowance constant*. The process is declared out-of-control if $C_n > h$, where the *control limit* $h$ is determined by setting the in-control "average run length" ($ARL$) at a certain nominal level $ARL_0$, and the in-control $ARL$ is defined to be the expected time to signal under $F$, that is,

$$(2) \qquad ARL = \mathbb{E}_F \inf\{n > 0 : C_n > h\}.$$

This is similar to setting the probability of Type I error at a specific level in the hypothesis testing context, with the null hypothesis being that the process is in control. If $\delta$ is the amount of shift in mean from $F$ to $G$, then choosing $k = \delta/2$ in (1.1) is optimal under certain regularity conditions [see, e.g., Reynolds (1975), Siegmund (1985)]. Similar CUSUMs exist in the literature for detecting downward or two-sided shifts in mean, or shifts in variance; see Hawkins and Olwell (1998).

An issue with the conventional CUSUM is its sensitivity to the assumption that both $F$ and $G$ are normal distributions with known in-control parameters. This fact is further confirmed by our numerical studies reported in Section 4. Depending on whether the true distribution $F$ is left or right skewed, whether it is heavy tailed or multimodal, the CUSUM may show two kinds of behavior. It may have very short or very long actual in-control $ARL$, compared to the nominal in-control $ARL$ value. In a hypothesis testing context, this is similar to the case that the actual probability of Type I error is larger or smaller than the nominal significance level of the test. The situation when the actual in-control $ARL$ value is much larger than the nominal value $ARL_0$ is clearly unacceptable in most applications, because in such cases a process that is actually out of control would not be detected as such for a long time. When the actual in-control $ARL$ value is much smaller than $ARL_0$, the CUSUM would be too sensitive to random noise, resulting in a large number of false alarms. Consequently, efficiency of the work related to the process being monitored would be negatively affected. The closeness of the actual in-control $ARL$ value to $ARL_0$ is related to the robustness of the CUSUM to various assumptions behind it, which is not a well studied topic in SPC.

In the absence of explicit knowledge of $F$, the bootstrap may be used for calibration of the control limit $h$, so that the actual in-control $ARL$ value matches the nominal value $ARL_0$. In the past 25 years or so, bootstrap techniques [see, e.g., Efron (1979), Efron and Tibshirani (1993), Shao and Tu (1995)] have been successful in obtaining highly accurate confidence intervals, estimates of asymptotic variances and other moments and probabilities, calibrations of different statistics, and so forth.

By the bootstrap technique, we draw repeated samples with replacement from observed data, and estimate the sampling distribution of a related statistic using these resamples. Implementation of the bootstrap is algorithmic; it often works under less stringent assumptions than classical asymptotics. In finite sample cases, since it uses the observed data efficiently by resampling, it can obtain more accurate results in many problems, compared to asymptotics-based classical techniques. For detailed discussion and examples, see Efron and Tibshirani (1993). However, the bootstrap is not always consistent; its properties depend on the problem and the statistic under consideration.

We performed a simulation experiment to study how well the bootstrap distribution of $C_n$ approximates its actual distribution. Before describing the simulation in Example 1.1 below, we introduce another statistic to facilitate the subsequent discussion. Let

$$(3) \quad T_n = \begin{cases} 0, & \text{if } C_n = 0, \\ j, & \text{if } C_n \neq 0, \ldots, C_{n-j+1} \neq 0, \ C_{n-j} = 0; \ j = 1, 2, \ldots, n. \end{cases}$$

Thus, $T_n$ is the time elapsed since the last time the CUSUM $C_n$ was zero, in view of which we call $T_n$ the *sprint length*. Note that $T_n$ can be computed easily, and $(C_n, T_n)$ forms a Markov process.

EXAMPLE 1.1.   We take $F$ to be the standard Normal distribution $N(0,1)$, $n = 10000$, and the allowance constant $k$ to be 0.5 which is optimal for detecting shifts from $F = N(0,1)$ to $G = N(1,1)$. We first approximate the distributions of $[C_n | T_n = j]$, for $j = 1, \ldots, 10$, by their empirical distributions, obtained from a preliminary run of 100,000 independent replications for each $j = 1, 2, \ldots, 10$. Then, for each $j$, another $I = 1000$ independent replications of sampling $X_1, \ldots, X_n$ from $F$ is performed. For each of these 1000 replications, we sample with replacement from the data and compute the CUSUM and sprint length statistics from the resample to obtain $B = 2000$ independent values of $[C_n^* | T_n^* = j]$. The empirical distribution function of these 2000 values is taken as the bootstrap estimate of the distribution of $[C_n | T_n = j]$. The $p$-value, corresponding to the Kolmogorov–Smirnov test for the null hypothesis that distributions of $[C_n | T_n = j]$ and $[C_n^* | T_n^* = j]$ are the same, is then computed for each of the 1000 replications. For each $j$, if the distribution of $[C_n^* | T_n^* = j]$ approximates the distribution of $[C_n | T_n = j]$ well, then the 1000 $p$-values computed above should roughly follow the Uniform distribution on $(0, 1)$. We perform another Kolmogorov–Smirnov test for testing whether this is true. Table 1 reports the $p$-value (in percent) of this second Kolmogorov–Smirnov test, for $j = 1, \ldots, 10$. From the table, it can be seen that the distribution of $[C_n^* | T_n^* = j]$ is a good approximation for that of $[C_n | T_n = j]$. This broad conclusion holds when this experiment is repeated with several other choices of $F$, $n$, $k$, $I$, $B$ and some other measures of closeness of distributions.

TABLE 1
*The p-value (in percentage) of the Kolmogorov–Smirnov test, for comparing the distribution of the p-value in testing equality of distributions of $[C_n | T_n = j]$ and $[C_n^* | T_n^* = j]$, with the Uniform distribution on $[0, 1]$*

| $j$ | 1 | 2 | 3 | 4 | 5 | 6 | 7 | 8 | 9 | 10 |
|---|---|---|---|---|---|---|---|---|---|---|
| $p$-value | 47.94 | 98.81 | 75.07 | 73.60 | 82.00 | 91.51 | 86.77 | 79.23 | 61.52 | 87.20 |

In this paper we propose a process control technique, using a sequence of control limits determined by the distributions of $[C_n | T_n = j]$, for different $j$. These control limits are obtained using bootstrap approximations, supported by the numerical study in Example 1.1. A motivation for considering the conditional distribution of $[C_n | T_n = j]$ in determining the control limits is that its distribution is much simpler to study, compared to the conventional, unconditional distribution of $C_n$. Under some regularity conditions, it can be checked that this conditional distribution depends only on $j$ and the in-control distribution $F$, but not on $n$. If $F$ is known, a recursive formula can be used to obtain the distributions of $[C_n | T_n = j]$, for $j = 1, \ldots$. In such cases, determination of control limits $\{h_j, j \geq 1\}$, computing powers of the related tests, and handling certain other statistical issues are just routine algebraic exercises. If $F$ is unknown, as is generally the case in applications, then from in-control data the distributions of $[C_n | T_n = j]$ can be approximated using the bootstrap, at least for relatively small values of $j$. In this paper we suggest estimating the control limits $h_j$'s using the bootstrap up to some value $j_{\max}$, after which a constant control limit is used.

Resampling techniques for SPC are of considerable recent interest in the literature. The bootstrap for assessing process capability was discussed by Franklin and Wasserman (1992). Shewhart charts based on the bootstrap are discussed by several authors, including Bajgier (1992), Seppala et al. (1995), Liu and Tang (1996) and Willemain and Runger (1996). Apart from the fact that some of these bootstrap methods are for Shewhart charts while we focus on the CUSUM, one major difference between these papers and ours is that they use the bootstrap mainly for estimating the distribution of the *run length*. In their methods the control limit $h$ is a constant and is chosen based on the assumption that $F$ is a known Normal distribution. When $F$ is unknown, or misspecified, their results will not be reliable. In this paper we use the bootstrap for approximating the distribution of the CUSUM statistic, conditional on $T_n$. We use these to obtain a sequence of control limits. Our procedure is distribution-free and robust to the specification of $F$. Wu and Wang (1996) and Wood, Kaye and Capon (1999) also design bootstrap-based control charts, though not for the CUSUM. Steiner (1999) suggests using time-varying control limits in the framework of exponentially weighted moving average (EWMA) control charts. In applications, a process may go out-of-control in a number of different ways; hence, controlling against broad alternatives is desirable. Our bootstrap-based SPC procedure is an attempt in that direction, and requires fewer and less restrictive assumptions than the conventional CUSUM. In particular, we may drop the assumptions that (a) $F$ is a Normal distribution, (b) $G$ is a Normal distribution, and (c) the in-control mean $\mu$ and the in-control standard deviation $\sigma$ are both known. Since our proposed method does not depend on $G$, it is robust against a variety of out-of-control situations. The bootstrap CUSUM presented in this paper is designed for detecting upward shifts in location parameters of $F$ only, which is similar to the conventional CUSUM (1). Other versions of the bootstrap CUSUM for detecting downward or two-sided shifts can be defined in a similar way.

In the literature efforts have been made to remove certain assumptions of the conventional CUSUM. For instance, Hawkins and Olwell (1998) suggested using the self-starting CUSUM when both $F$ and $G$ are Normal but the in-control distribution parameters are unknown. Several nonparametric CUSUMs have also been proposed. See Chakraborti, van der Laan and Bakir (2001) for a review of 1-dimensional methods, and Qiu (2008) and Qiu and Hawkins (2001, 2003) for multivariate nonparametric CUSUMs. In Section 3 we describe a nonparametric CUSUM by Bakir and Reynolds (1975) that we use for comparison with our method. In Section 4 we consider some illustrative numerical examples where the conventional CUSUM, the nonparametric CUSUM and the proposed bootstrap CUSUM are compared. In Section 5 we apply these techniques to a real-data problem in the aluminum smeltering industry. In Section 6 we briefly summarize our conclusions for this study.

**2. The proposed SPC procedure.** Statistical process control has two phases. In Phase I a set of process data is gathered and analyzed. Any 'patterns' in this data-set indicating a lack of statistical control would lead to adjustments and fine tuning of the process. Once all such process calibration issues are addressed, a set of clean data is obtained, gathered under stable operating conditions and illustrative of the actual process performance. The techniques discussed in this paper are for Phase II SPC, where the process is monitored to detect possible out-of-control behaviors. In our study, we make use of the clean, in-control, Phase I data to set up our proposed CUSUM procedure, as discussed below.

For simplicity of discussion, let $Y_j$ be a random variable having the distribution of $[C_n|T_n = j]$. For any positive integer $j_{\max} \leq n$, the distribution of $C_n$ equals that of $\sum_{j=1}^{j_{\max}} Y_j I_{\{T_n=j\}} + Y^* I_{\{T_n > j_{\max}\}}$, where $Y^*$ is a random variable with the distribution of $[C_n|T_n > j_{\max}]$. Hence, when $T_n = j$ $(T_n > j_{\max})$, it is reasonable to choose the control limit $h_j$ $(h^*)$ based on the distribution of $Y_j$ $(Y^*)$. At time point $n$, the process is declared to be out-of-control if $T_n = j$ and $C_n > h_j$, for $1 \leq j \leq j_{\max}$, or if $T_n > j_{\max}$ and $C_n > h^*$. The constants $j_{\max}$ and $k$ are the two tuning parameters of this procedure, whose choice is up to the practitioner. Since the control limits $\{h_j, 1 \leq j \leq j_{\max}; h^*\}$ are obtained using the bootstrap, the choice of $j_{\max}$ is limited by the in-control data, allowance constant $k$, and available computational power. Our first step is to fix $j_{\max}$ as large as is convenient based on computational considerations. Our simulations described in Section 4 show that a high $j_{\max}$ value need not result in the most efficient bootstrap based SPC, for a given pair of in and out-of-control distributions $F$ and $G$. In the simulation problems we investigated, the results seem to be fairly stable for $j_{\max}$ in the range 20–50, but not necessarily so for smaller values of $j_{\max}$.

Note that owing to the nonlinear and nonsmooth nature of $C_n$ in (1), the control limits $\{h_j, 1 \leq j \leq j_{\max}; h^*\}$ and the distribution of $T_n$ are intractable functions of the allowance constant $k$. In conventional CUSUMs, selection of $k$ is related to $\delta$, the shift size in the mean of $F$. Since we desire robustness against non-Normality and the value of $\delta$ for our SPC method, we suggest selecting $k$ based on the average sprint length $\mathbb{E}T_n$. From expressions (1) and (3), it can be seen that, if $k$ is chosen larger, then $C_n$ will have a larger chance to bounce back to 0. Consequently, $\mathbb{E}T_n$ would be smaller. Similarly, if $k$ is chosen smaller, then $\mathbb{E}T_n$ would be larger. In the SPC literature it is already well demonstrated that larger $k$ values are good for detecting larger shifts, and vice versa. Therefore, selection of $\mathbb{E}T_n$ should be an important issue. In our numerical examples presented in Sections 4 and 5, we consider three choices for $\mathbb{E}T_n$, namely, $\mathbb{E}T_n = 0.5j_{\max}$, $\mathbb{E}T_n = 0.75j_{\max}$, and $\mathbb{E}T_n = j_{\max}$, to represent small, moderate, and large values of $\mathbb{E}T_n$.

Obtaining the value of $k$ from $\mathbb{E}T_n$ is a simple iterative computation, described briefly below. Let $k_L$, $k_U$, and $k_0$ be the lower-bound, upper-bound, and an initial value of $k$. Draw $B$ bootstrap samples from the normalized in-control data (i.e., having zero sample mean and unit sample variance). In the first iteration, the CUSUM procedure uses allowance constant $k_0$. Based on each bootstrap sample, we record the value of the first sprint length of the CUSUM; thus, $B$ values of the sprint length can be recorded from the $B$ bootstrap samples. Then, $\mathbb{E}T_n$ is estimated by the sample mean of these $B$ sprint length values. If the estimated $\mathbb{E}T_n$ is larger than the target $\mathbb{E}T_n$ value, then we update $k$ to be $k_1 = (k_U + k_0)/2$, and use $k_0$ and $k_U$ as the new lower and upper bounds. Otherwise, update $k$ to be $k_1 = (k_L + k_0)/2$, and use $k_L$ and $k_0$ as the new lower and upper bounds. Go to the next iteration after replacing $k_0$ by $k_1$. This process continues until the estimated $\mathbb{E}T_n$ value in an iteration is close enough to the target $\mathbb{E}T_n$ value. In all our numerical examples presented in Sections 4 and 5, we take $k_L$, $k_U$, and $k_0$ to be the first, third, and second quartiles of the in-control data, and $B = 5000$. The above binary search procedure converges very fast, taking about 10 iterations to achieve the pre-specified accuracy.

Once $j_{\max}$ and $k$ are fixed, the sequence of control limits $\{h_j, j \leq j_{\max}; h^*\}$ can be determined from the in-control data using the bootstrap. These control limits are related to certain tail probabilities of the in-control distribution $F$. In the literature it has been demonstrated that, in such cases, it is better to first estimate $F$ by $\hat{F}$ using a density estimation procedure and then obtain the bootstrap samples from $\hat{F}$ (i.e., using the *smoothed bootstrap*), compared to drawing bootstrap samples directly from the observed data [cf., e.g., Hall, DiCiccio and Romano (1989), Falk and Reiss (1989)]. In this paper we construct a kernel smoothing density estimator for the Phase I data, and then take the corresponding distribution as $\hat{F}$. The bandwidth used in kernel smoothing is chosen by cross validation. Bootstrap samples are then drawn from $\hat{F}$, using a procedure described in Silverman (1986).

Our algorithm for determining $\{h_j, j \leq j_{\max}; h^*\}$ consists of two steps. In the first step the bootstrap is used for obtaining preliminary values $\{M_j, j \leq j_{\max}, M^*\}$, such that $M_j \approx h_j$ and $M^* \approx h^*$. Then, in the second step these values are calibrated using some more bootstrap steps to ensure that the resulting in-control average run length, denoted as $ARL$, equals the nominal $ARL_0$ up to a certain level of accuracy.

The following algorithm describes how to obtain $\{M_j, j \leq j_{\max}, M^*\}$. Let $B$ be the bootstrap Monte Carlo sample size, $C^*_{\text{old}} = 0$, $T^*_{\text{old}} = 0$, and $b = 0$. For all $j \in \{1, \ldots, j_{\max} + 1\}$, we implement the following:

Step 0. Set $b = b + 1$.

Step 1. Draw an observation $X^*$ from $\hat{F}$.

Step 2. Update $C^*_{\text{old}}$ to $C^*_{\text{new}} = \max(C^*_{\text{old}} + X^* - k, 0)$. If $C^*_{\text{new}} > 0$, then compute $T^*_{\text{new}}$ by $T^*_{\text{new}} = T^*_{\text{old}} + 1$. If $C^*_{\text{new}} = 0$, then set $T^*_{\text{new}} = 0$.

Step 3. Check if $T^*_{\text{new}} = j$. If so, then record $Y_{j:b} = C^*_{\text{new}}$. If not, then set $C^*_{\text{old}} = C^*_{\text{new}}$ and $T^*_{\text{old}} = T^*_{\text{new}}$, and go to Step 1. If $b < B$, go to Step 0.

At the end of an execution of this algorithm, we would have $B$ numbers $Y_{j:1}, Y_{j:2}, \ldots, Y_{j:B}$. Define

$$(4) \qquad \widehat{\alpha} = (\widehat{p}^2 ARL_0)^{-1},$$

where $\widehat{p}$ denotes the proportion of observations in the in-control data that are larger than $k$. Then, the $B(1-\widehat{\alpha})$th ordered value from $Y_{j:1}, Y_{j:2}, \ldots, Y_{j:B}$ is taken as $M_j$, for $j \leq j_{\max}$. The $B(1-\alpha)$th ordered value from $Y_{(j_{\max}+1):1}, \ldots, Y_{(j_{\max}+1):B}$ is taken as $M^*$. The formula (4) is based on some asymptotic approximations.

Next, we describe the algorithm to fine tune $M_1, \ldots, M_{j_{\max}}$ and $M^*$ to obtain $h_1, \ldots, h_{j_{\max}}$ and $h^*$ so that the nominal $ARL_0$ is reached. This algorithm is iterative. In the first iteration, define $h_j^{(0)} = M_j$, for $1 \leq j \leq j_{\max}$, and $h^{*(0)} = M^*$. Let $C_0 = 0$, and $T_0 = 0$. For $n \geq 1$, generate $X_n$ from $\widehat{F}$, construct $C_n = \max\{C_{n-1} + X_n - k, 0\}$, and keep track of the corresponding sprint length $T_n$. If $T_n = j$ and $C_n > h_j^{(0)}$, then declare the process to be out-of-control and take the run length as $n$. If $T_n > j_{\max}$ and $C_n > h^{*(0)}$, we also declare the process to be out-of-control and take the run length as $n$. Repeat this $N_1$ times (e.g., $N_1 = 100$ in our numerical examples reported in Section 4) and define the average of these $N_1$ run lengths as $RL^{(0)}$. If $RL^{(0)} < ARL_0$, then repeat the above procedure, after $h_1^{(0)}, \ldots, h_{j_{\max}}^{(0)}$ and $h^{*(0)}$ are replaced by $h_{1U}^{(0)} = (1+\varepsilon)M_1, \ldots, h_{j_{\max}U}^{(0)} = (1+\varepsilon)M_{j_{\max}}$ and $h_U^{*(0)} = (1+\varepsilon)M^*$, where $\varepsilon > 0$ is a parameter. The corresponding averaged run length is denoted by $RL_U^{(0)}$. Define

$$h_j^{(1)} = \frac{RL_U^{(0)} - ARL_0}{RL_U^{(0)} - RL^{(0)}} h_j^{(0)} + \frac{ARL_0 - RL^{(0)}}{RL_U^{(0)} - RL^{(0)}} h_{jU}^{(0)}, \qquad \text{for } j = 1, \ldots, j_{\max},$$

$$(5)$$

$$h^{*(1)} = \frac{RL_U^{(0)} - ARL_0}{RL_U^{(0)} - RL^{(0)}} h^{*(0)} + \frac{ARL_0 - RL^{(0)}}{RL_U^{(0)} - RL^{(0)}} h_U^{*(0)}.$$

Also, define $h_{jU}^{(1)} = h_j^{(0)}$, $h_{jL}^{(1)} = h_j^{(0)}$ for $j = 1, \ldots, j_{\max}$, and $h_U^{*(1)} = h_U^{*(0)}$, $h_L^{*(1)} = h^{*(0)}$. If $RL^{(0)} > ARL_0$, then run the CUSUM procedure using control limits $h_{1L}^{(0)} = (1-\varepsilon)M_1, \ldots, h_{j_{\max}L}^{(0)} = (1-\varepsilon)M_{j_{\max}}$ and $h_L^{*(0)} = (1-\varepsilon)M^*$; the corresponding averaged run length is denoted as $RL_L^{(0)}$. In this case, $\{h_j^{(1)}, 1 \leq j \leq j_{\max}, h^{*(1)}\}$ are defined similarly to those in (5), as linear interpolations of $\{h_j^{(0)}, 1 \leq j \leq j_{\max}, h^{*(0)}\}$ and $\{h_{jL}^{(0)}, 1 \leq j \leq j_{\max}, h_L^{*(0)}\}$, with weights $(ARL_0 - RL_L^{(0)})/(RL^{(0)} - RL_L^{(0)})$ and $(RL^{(0)} - ARL_0)/(RL^{(0)} - RL_L^{(0)})$, respectively. Further, we define $h_{jU}^{(1)} = h_j^{(0)}$, $h_{jL}^{(1)} = h_{jL}^{(0)}$ for $j = 1, \ldots, j_{\max}$ and $h_U^{*(1)} = h^{*(0)}$, $h_L^{*(1)} = h_L^{*(0)}$.

The parameter $\varepsilon$ should be chosen such that $RL_U^{(0)} > ARL_0$ and $RL_L^{(0)} < ARL_0$. The second iteration is the same as the first iteration, except that $\{h_j^{(0)}, 1 \le j \le j_{\max}, h^{*(0)}\}$, $\{h_{jL}^{(0)}, 1 \le j \le j_{\max}, h_L^{*(0)}\}$, and $\{h_{jU}^{(0)}, 1 \le j \le j_{\max}, h_U^{*(0)}\}$ need to be replaced by $\{h_j^{(1)}, 1 \le j \le j_{\max}, h^{*(1)}\}$, $\{h_{jL}^{(1)}, 1 \le j \le j_{\max}, h_L^{*(1)}\}$, and $\{h_{jU}^{(1)}, 1 \le j \le j_{\max}, h_U^{*(1)}\}$. At the end of this iteration, we obtain $\{h_j^{(2)}, 1 \le j \le j_{\max}, h^{*(2)}\}$, $\{h_{jL}^{(2)}, 1 \le j \le j_{\max}, h_L^{*(2)}\}$, and $\{h_{jU}^{(2)}, 1 \le j \le j_{\max}, h_U^{*(2)}\}$, similar to those in the first iteration. This iterative algorithm continues until the $k$th iteration in which the exit condition $|RL^{(k)} - ARL_0|/ARL_0 < \tilde{\varepsilon}$ is satisfied, where $\tilde{\varepsilon} > 0$ is a pre-specified small number. In our simulations reported in the next section, we took $\varepsilon = 0.2$ and $\tilde{\varepsilon} = 0.02$, and found that usually less than 5 iterations were required to satisfy the exit condition.

Note that in the above procedure $k$ is linked to $j_{\max}$, which was not chosen to be optimal for the problem at hand. A practitioner may skip the step of adaptively choosing $k$ and use a fixed constant instead. However, our simulations (not reported here) suggest the above method of choosing $k$ adaptively leads to better performance than using a fixed $k \in [0, 1]$. Another reason for linking $ET_n$, and hence $k$, to $j_{\max}$ is that the probabilities of the events $\{T_n = j\}$ decrease sharply with increase of either $k$ or $j$. So beyond a data-dependent range of $k$ and $j$ values, the probability of observing the events $(T_{\text{new}}^* = j)$ in the above algorithm is essentially zero. For instance, consider the scenario when Step 1 of the algorithm is implemented by drawing $X^*$ randomly from the observed data $X_1, \ldots, X_m$, instead of from the smoothed density $\hat{F}$. In the extreme case when $k > \max_{1 \le i \le m} X_i$, $C_n^*$ does not have any positive jumps. Consequently, none of the $h_j$'s can be estimated. If $k$ is between the top two order statistics of $X_1, \ldots, X_m$, $C_n^*$ can increase only in steps of $\max_{1 \le i \le m} X_i - k$, and the estimates of the $h_j$ would reflect this nonsmoothness. In addition to using smoothed bootstrap and calibrating $\{M_j, j \le j_{\max}; M^*\}$, using a relatively small $k$, or equivalently, a relatively large $ET_n$ and $j_{\max}$, would help in obtaining better estimates of $\{h_j, j \le j_{\max}; h^*\}$. On the other hand, using large $k$ helps in quickly detecting large shifts from $F$.

If $k$ is taken to be a fixed constant, and one wants to reduce the number of control limits used from $j_{\max} + 1$ to $\tilde{j}_{\max} + 1$, one option is to leave the first $\tilde{j}_{\max}$ control limits unchanged and recompute $h^*$ only using the above calibration step. This may be used even when $k$ is adaptively chosen, but in that case the $\mathbb{E}T_n/j_{\max}$ ratio is no longer preserved.

Let us list some shortcomings of the proposed method here. First, since its construction does not depend on the out-of-control distribution $G$, it is expected to be less sensitive to any specific choice of $G$, compared to a conventional CUSUM using certain prior information about $G$. For instance, in cases where both $F$ and $G$ are Normal and the in-control mean and variance are known, the conventional CUSUM (1) would outperform our proposed procedure. Second, our method depends on how closely $\hat{F}$ approximates $F$. We use a kernel smoothed density estimator in this paper, owing to its simplicity. Other choices of nonparametric distribution or density estimation may also be used, although an in-control phase I data of moderate size is always required. Third, it requires considerable computation in setting up our method. For instance, in a typical case considered in our numerical examples in Section 4, it requires about 20 seconds computing time to determine the values of $k$ and $\{h_j, 1 \le j \le j_{\max}; h^*\}$ on our dual-processor Pentium III PC with 800 MHz CPU. However, once these values are obtained, the monitoring process is just routine.

**3. Description of an existing nonparametric CUSUM.** Some attempts have been made in the literature to rectify the conventional CUSUM by overcoming some of its obvious deficiencies. As mentioned in Section 1, there is a large body of literature that attempts to substitute rank- and sign-based statistics in place of the original observations. Among those existing nonparametric control charts, the one by Bakir and Reynolds (1979) is classical and often used as a gold standard in the nonparametric control charts literature [cf., e.g., Chakraborti, van der Laan and Bakir (2001)]. In this section we briefly introduce this procedure. We will make some numerical comparisons between our proposed method and this method in Section 4.

As in conventional phase II SPC, the mean of the in-control measurement distribution $F$ is assumed known. Without loss of generality, it is assumed to be 0. By Bakir and Reynolds's method, the observed data is grouped into blocks of size $g$ each. Then, define $R_{ij}$ as the rank of the absolute value of the $j$th observation in the $i$th block. That is, $R_{ij}$ is the rank of $|X_{ij}|$ among $\{|X_{i1}|, |X_{i2}|, \ldots, |X_{ig}|\}$, where $i = 1, 2 \ldots$. Then define $U_{ij} = \text{sign}(X_{ij})R_{ij}$ and $V_i = \sum_{j=1}^{g} U_{ij}$, and construct a CUSUM based on the $V_i$. That is, we look at the process $S_n = \max(0, S_{n-1} + V_n - k)$. The process being monitored is declared to be out-of-control if $S_n > h$.

A crucial assumption of this procedure is that $F$ is a symmetric distribution. There are several other features to be noted for this nonparametric CUSUM. First, for fixed $k$, the distribution of $S_n$ does not depend on that of $X_1$, hence, it is distribution-free. Second, by replacing the observations by their within-group ranks, it appropriately scales for outliers. Third, since the procedure requires a grouping, results may depend on the value of $g$ selected. Fourth, the $V_i$'s take integer values. While this ensures a certain amount of insensitivity to chance errors, it also implies that there may not be a $h$ value corresponding to a given value of $ARL_0$. In fact, Table 2a of Bakir and Reynolds (1979) confirms this fact rather dramatically, which presents some consecutive $h$ values and the corresponding $ARL_0$ values for certain fixed $g$ and $k$, and the $ARL_0$ values are widely apart. Fifth, if the parameter $k$ is to be chosen optimally, then a knowledge of $F$ is essential, which obviously defeats the purpose of having a distribution-free procedure. Sixth, the procedure is insensitive to shifts in the scale parameter, but it is sensitive to shifts in the location parameter. In many cases, it is unknown whether the shift in $F$ is in its location parameter only. Finally, there does not seem to be a satisfactory procedure for computing $h$. Indeed, $h$ appears to differ in cases with different $F$ (and $G$), which would also defeat the purpose of having a distribution-free procedure.

**4. Simulation studies.** In this section we present some numerical examples for investigating the performance of the proposed procedure. Recall that there are two major ideas in the proposed procedure: (i) using a set of control limits $\{h_1, \ldots, h_{j_{\max}}, h^*\}$, instead of a single control limit, and (ii) using the bootstrap for estimating these control limits. In order to study the two ideas separately, we first assume that the in-control distribution $F$ is known. In such cases, $\{h_1, \ldots, h_{j_{\max}}, h^*\}$ can be computed directly from $F$; thus, bootstrapping is not needed. When $F$ is known, the conventional CUSUM is optimal if $F$ is a normal distribution and its mean and variance are both known. However, when $F$ is not a normal distribution, results from the conventional CUSUM could be misleading, since its true in-control average run length could be very different from the nominal one $ARL_0$. The proposed procedure, on the other hand, is still reliable in such cases, because its control limits are computed from $F$. To see these facts, we consider the following three cases:

Case I $F = N(0, 1)$ and $G = N(\delta, 1)$.

Case II $F$ has the density function $(1/6)\exp(-x/3)$ when $x \geq 0$ and $(1/2)\exp(x)$ when $x < 0$, standardized to have mean 0 and variance 1. $G$ is a location shift of $F$ with shift size $\delta$. In this case, $F$ is skewed to the right.

Case III $F$ has the density $(1/6)\exp(x/3)$ when $x < 0$ and $(1/2)\exp(-x)$ when $x \geq 0$, standardized to have mean 0 and variance 1. $G$ is a location shift of $F$ with shift size $\delta$. In such cases, $F$ is skewed to the left.

In the first part of the simulations we use the algorithm presented in Section 2 to generate our SPC procedure (denoted as **B**), but with $F$ in place of $\hat{F}$, since $F$ is known. In the algorithm we choose $B = 5000$, $j_{\max} = 50$, and $\mathbb{E}T_n = 37.5$. Other parameters are chosen as those specified in Section 2. In all three cases we take $\delta$ to be 0 or 0.5, and $ARL_0 = 200$. We assume that $\delta$ is known when implementing the classical CUSUM (denoted as **C**), and is unknown when implementing our proposed CUSUM, hence, we bias the results in favor of the classical CUSUM. For the classical CUSUM, we take the allowance constant to be $k = 0.25$. The in- and out-of-control average run lengths over 1000 replications, along with the standard errors of the average run lengths, of the two procedures are presented in Table 2.

From Table 2, it can be seen that, in case I when the normal assumption is valid and when the in-control mean and variance are known, the classical CUSUM **C** performs well. Its actual in-control $ARL$ is close to 200. The expected time to detection after the process is out-of-control, denoted by $ARL_1$, is relatively small, as expected. In such a case, the proposed procedure **B** is comparable. However, in cases II and III when $F$ is skewed to the right or left, the actual in-control $ARL$ values of **C** are well above or below 200. In case II **C** would not detect a potential shift as quick as we would expect, and it would provide a false signal of a process change with a larger than expected probability in case III. In comparison, the actual in-control $ARL$ values of procedure **B** are not significantly different from 200 (e.g., in case II, its estimated in-control $ARL$ value 207.11 is within 1 standard error of 200), and its $ARL_1$ values are reasonably small in these cases.

The above example shows that the classical CUSUM is sensitive to distributional assumptions. Our other numerical studies, which are not reported here, suggest that it is also sensitive to the choice of the tuning parameters like the allowance constant $k$ when $F$ is not Normal, and to the variability in estimates of the mean and variance of $F$ obtained from the Phase I data [see related discussion in Jones, Champ and Rigdon (2004)]. In comparison, our proposed procedure does not require prior information about both $F$ and $G$; consequently, it is robust to distributional assumptions. The above example also suggests that our method is competitive to the optimal classi-cal CUSUM in the Normal case (i.e., case I), while it performs more reliably when the normality assumption does not hold (i.e., cases II and III).

TABLE 2
*ARL values and their standard errors (in parentheses) of the classical CUSUM* **C** *and the proposed CUSUM* **B** *in cases* I–III. *The nominal $ARL_0$ is 200 in all cases*

|  | **C** | | **B** | |
|---|---|---|---|---|
| **Case** | $\delta = 0$ | $\delta = 0.5$ | $\delta = 0$ | $\delta = 0.5$ |
| I | 202.49 | 19.66 | 201.16 | 19.13 |
|  | (7.72) | (0.32) | (6.19) | (0.47) |
| II | 669.67 | 16.69 | 207.11 | 10.38 |
|  | (30.82) | (0.30) | (8.32) | (0.54) |
| III | 119.84 | 22.62 | 194.79 | 31.43 |
|  | (4.40) | (0.48) | (2.98) | (0.88) |

Next, we consider the three cases described above without assuming $F$ to be known for our algorithm. We study the properties of six different procedures. They are (i) the classical CUSUM (labeled as **C**), (ii) the nonparametric CUSUM by Bakir and Reynolds (1979) when its parameters $(g, k, h)$ are chosen to be (10, 13, 24) (labeled as **NP1**), (iii) the nonparametric CUSUM by Bakir and Reynolds when $(g, k, h)$ are chosen to be (10, 21, 14) (labeled as **NP2**), and (iv)–(vi) the bootstrap CUSUMs with allowance constants resulting from setting $\mathbb{E}T_n = 0.5j_{\max}$, $\mathbb{E}T_n = 0.75j_{\max}$, and $\mathbb{E}T_n = j_{\max}$, respectively (labeled as **B1**, **B2**, and **B3**). By using the relationship between $k$ and $\mathbb{E}T_n$, as discussed in Section 2, the corresponding $k$ values are respectively 0.028, 0.017, and 0.011 in procedures **B1**, **B2**, and **B3** when $j_{\max} = 50$ in case I.

Let the location shift from $F$ to $G$ be denoted as $\delta$. For **C**, we assume that $\delta$ is known, and we use the optimal allowance constant $k = \delta/2$. Its control limit $h$ is computed using a standard algorithm [cf. Hawkins and Olwell (1998), Chapter 2]. Regarding the nonparametric CUSUM by Bakir and Reynolds, there seems to be no simple algorithm for choosing its parameters $(g, k, h)$ in a distribution-free fashion, which makes it relatively inconvenient to use for many applications. Bakir and Reynolds provide several tables listing in-control $ARL$ values of their procedure for many different combinations of $g, k$, and $h$. The two cases considered (i.e., labeled **NP1** and **NP2** in this paper) have $ARL$ values close to 200. For procedures **B1**, **B2**, and **B3**, parameters other than $\mathbb{E}T_n$ and $j_{\max}$ are chosen as in the previous example.

We studied these six procedures in various situations when the nominal $ARL_0$ value is fixed at 200. First, $\delta$ is allowed to take the values 0, 0.50, and 1. The case with $\delta = 0$ is for studying the actual in-control $ARL$ values of the related SPC techniques. Second, in procedures **B1**, **B2**, and **B3**, $j_{\max}$ takes the value of 5, 30, 40, or 50, which allows us to investigate the possible effect of $j_{\max}$ on the performance of the proposed bootstrap procedures. For each simulation, phase I data of size $m$ is generated from $F$, and phase II data is generated sequentially, with the first $n_1$ observations from $F$ and the rest from $G$. In this paper we fix $m = 1000$ and $n_1 = 0$. The estimated mean and variance from the Phase I data are used in **C**. For the proposed bootstrap procedures, we found that results are already quite satisfactory when $m$ takes a value of about 100, although their performance improves with larger $m$. On the other hand **C** is extremely variable when $m$ is small, due to the variability induced by the estimated mean and variance from the Phase I data that are used for Phase II SPC.

In the $i$th simulation, we check the time point $RL_{ij}$ when the $j$th CUSUM technique sends an out-of-control signal. For the $j$th CUSUM technique, we report the sample mean $ARL_j = I^{-1} \sum_{i=1}^{I} R_{ij}$ and the associated standard error $SERL_j = I^{-1/2}\sqrt{\frac{1}{I-1}\sum_{i=1}^{I}(RL_{ij} - ARL_j)^2}$ of all $RL_{ij}$ values obtained from $I = 100$ simulations. The standard error $SERL_j$ gives us an idea of the variability associated with $ARL_j$. Simulation results in cases I–III are presented in Tables 3–5, respectively.

From Table 3, it can be seen that, in case I when the normality holds, the actual in-control $ARL$ values of **C**, and **B1**, **B2**, and **B3** when $j_{\max} = 30, 40$, or 50 are all within 1 standard error of the nominal $ARL_0$ value of 200. The nonparametric procedures **NP1** and **NP2** register low actual in-control $ARL$ values. By comparing different procedures with respect to their $ARL_1$ values, it can be seen that the optimal classical CUSUM generally performs well for all different $\delta$ values, and the bootstrap procedures are just as good. In fact, in some cases (e.g., in cases when $\delta = 1$ and $j_{\max} = 30, 40$, or 50), the bootstrap procedures out-perform **C**, although some differences among the related $ARL_1$ values are not large enough to be significant at the 0.05 significance level. This can be explained by the facts that the optimality of **C** is based on asymptotic theory [Lorden (1971)] and in the sense of minimizing $\max_{n_1 \geq 0} ARL_1$ where $n_1$ is the true shift time [Moustakides (1986)], while $m$ and $n_1$ are fixed in the current example. For the three bootstrap procedures, it seems that their performance becomes quite stable when $j_{\max} \geq 30$. When $j_{\max}$ is small (i.e., $j_{\max} = 5$), their performance may not be stable, in the sense that their actual $ARL$ values could be quite different from $ARL_0$ and their $ARL_1$ could be relatively large. Regarding the two **NP** procedures,

Table 3

*Average run lengths and their standard errors (in parentheses) of different SPC procedures in case* I. *Nominal $ARL_0$ value is* 200

|  |  | $\delta = 0$ | $\delta = 0.5$ | $\delta = 1$ |
|---|---|---|---|---|
|  | **C** | 206.96 (25.86) | 19.08 (1.39) | 7.53 (0.47) |
|  | **NP1** | 140.30 (13.49) | 21.40 (1.23) | 11.90 (0.42) |
|  | **NP2** | 155.20 (16.67) | 20.90 (1.32) | 11.60 (0.40) |
| $j_{\max} = 5$ | **B1** | 178.43 (20.03) | 25.31 (0.97) | 12.54 (0.57) |
|  | **B2** | 173.78 (20.16) | 18.37 (1.17) | 7.94 (0.47) |
|  | **B3** | 201.86 (24.73) | 27.38 (2.14) | 9.14 (0.87) |
| $j_{\max} = 30$ | **B1** | 202.92 (18.96) | 18.89 (1.45) | 6.60 (0.46) |
|  | **B2** | 194.44 (18.86) | 18.68 (1.49) | 6.43 (0.42) |
|  | **B3** | 197.79 (17.81) | 19.20 (1.56) | 6.45 (0.42) |
| $j_{\max} = 40$ | **B1** | 195.04 (21.04) | 22.40 (1.72) | 5.66 (0.46) |
|  | **B2** | 198.98 (21.19) | 20.52 (1.80) | 5.70 (0.48) |
|  | **B3** | 201.87 (22.66) | 21.36 (1.80) | 5.77 (0.48) |
| $j_{\max} = 50$ | **B1** | 190.88 (25.66) | 16.96 (1.61) | 6.59 (0.59) |
|  | **B2** | 199.35 (25.74) | 18.73 (1.62) | 6.84 (0.59) |
|  | **B3** | 202.79 (28.53) | 17.51 (1.60) | 6.50 (0.54) |

TABLE 4
*Average run lengths and their standard errors (in parentheses) of different SPC techniques in case* II. *Nominal ARL$_0$ value is 200*

|            |        | $\delta = 0$    | $\delta = 0.5$   | $\delta = 1$   |
|------------|--------|-----------------|------------------|----------------|
|            | **C**  | 258.96 (28.76)  | 78.04 (7.45)     | 36.42 (3.25)   |
|            | **NP1**| 656.90 (67.97)  | 176.30 (16.12)   | 57.10 (4.57)   |
|            | **NP2**| 635.90 (60.52)  | 173.30 (15.15)   | 75.30 (6.99)   |
| $j_{\max} = 5$  | **B1** | 191.13 (18.55)  | 62.65 (3.50)     | 37.58 (1.57)   |
|            | **B2** | 191.86 (19.32)  | 58.09 (4.03)     | 23.27 (1.53)   |
|            | **B3** | 204.71 (20.22)  | 93.38 (9.68)     | 34.98 (3.30)   |
| $j_{\max} = 30$ | **B1** | 205.59 (21.63)  | 103.37 (8.66)    | 48.91 (2.80)   |
|            | **B2** | 204.24 (22.03)  | 100.32 (7.65)    | 48.38 (2.61)   |
|            | **B3** | 207.31 (25.49)  | 99.21 (7.76)     | 48.21 (2.70)   |
| $j_{\max} = 40$ | **B1** | 203.95 (25.65)  | 112.21 (9.09)    | 52.72 (4.10)   |
|            | **B2** | 206.69 (24.74)  | 113.57 (9.11)    | 52.77 (4.55)   |
|            | **B3** | 204.78 (23.22)  | 106.66 (7.87)    | 52.74 (4.55)   |
| $j_{\max} = 50$ | **B1** | 195.90 (19.85)  | 102.74 (8.19)    | 53.27 (3.93)   |
|            | **B2** | 195.42 (21.74)  | 98.09 (7.73)     | 53.55 (3.94)   |
|            | **B3** | 194.02 (20.83)  | 98.63 (8.16)     | 53.05 (3.88)   |

TABLE 5
*Average run lengths and their standard errors (in parentheses) of different SPC techniques in case* III. *Nominal ARL$_0$ value is 200*

|            |        | $\delta = 0$    | $\delta = 0.5$   | $\delta = 1$    |
|------------|--------|-----------------|------------------|-----------------|
|            | **C**  | 232.61 (24.16)  | 82.74 (7.64)     | 57.25 (4.74)    |
|            | **NP1**| 51.50 (4.62)    | 31.90 (2.33)     | 26.10 (1.70)    |
|            | **NP2**| 55.20 (4.85)    | 33.70 (2.66)     | 25.20 (1.69)    |
| $j_{\max} = 5$  | **B1** | 199.38 (18.47)  | 73.38 (5.06)     | 40.09 (1.72)    |
|            | **B2** | 204.2 (26.4)    | 75.07 (13.46)    | 64.48 (5.16)    |
|            | **B3** | 216.17 (27.05)  | 116.85 (16.46)   | 79.67 (16.55)   |
| $j_{\max} = 30$ | **B1** | 199.51 (19.45)  | 54.59 (3.66)     | 28.17 (2.57)    |
|            | **B2** | 195.73 (21.69)  | 54.33 (3.66)     | 28.17 (2.55)    |
|            | **B3** | 190.03 (19.38)  | 55.34 (3.59)     | 28.59 (2.49)    |
| $j_{\max} = 40$ | **B1** | 202.34 (21.54)  | 49.62 (4.27)     | 22.67 (2.50)    |
|            | **B2** | 197.50 (21.50)  | 48.33 (4.33)     | 22.45 (2.48)    |
|            | **B3** | 184.37 (16.45)  | 48.22 (4.21)     | 21.91 (2.34)    |
| $j_{\max} = 50$ | **B1** | 197.15 (18.51)  | 45.53 (4.87)     | 18.93 (2.26)    |
|            | **B2** | 190.68 (20.61)  | 49.59 (5.17)     | 19.85 (2.35)    |
|            | **B3** | 193.50 (19.54)  | 50.88 (5.19)     | 17.50 (2.24)    |

although their actual in-control $ARL$ values are much less than 200, their $ARL_1$ values are quite large especially when $\delta$ is large (i.e., $\delta = 1$). Therefore, in this case, they are not as efficient as the other methods considered here for phase II SPC, which is expected because of the loss of information by using ranks and grouped data.

From Table 4, it can be seen that, when $F$ is right skewed, the conventional CUSUM **C** has a quite large actual in-control $ARL$ value. The two nonparametric procedures **NP1** and **NP2** are misleading in this case because their actual in-control $ARL$ and $ARL_1$ values are all very large. In comparison, the proposed bootstrap CUSUMs perform reasonably well. Their actual in-control $ARL$ values are within 1 standard error of 200 in all cases. It is worth mentioning that, in this case, the proposed bootstrap CUSUMs seem to perform better when $j_{max}$ is chosen smaller, especially when $\delta = 1$. This might be because, by our algorithm, the choice of a small $j_{max}$ results in large $k$, which makes detection of shifts easier. This requires further investigation.

From Table 5, if $F$ is left skewed, the nonparametric procedures register extremely small actual in-control $ARL$ values. It can be seen that the actual in-control $ARL$ value of **C** is larger than 200, although their difference is only about 1.4 times the standard error. The $ARL_1$ values of **C** are large in both cases when $\delta = 0.5$ and $\delta = 1$. In comparison, the bootstrap procedures perform well, except when $j_{max} = 5$. From the table, the bootstrap procedures seem to stabilize when they are used for detecting the smaller shift $\delta = 0.5$ and when $j_{max}$ is chosen as large as 50. When it is used for detecting the larger shift $\delta = 1$, it seems that performance of the bootstrap procedures still has room for improvement by using larger $j_{max}$ values. This suggests that optimal selection of $j_{max}$ may depend not only on the shape of $F$, but also on the shift size.

It should be pointed out that $ARL$ values from different procedures acting on the same data are usually positively associated. Therefore, when we compare two different methods based on their $ARL_1$ values listed in the above tables, the actual $p$-value would be less than the one obtained when they are assumed independent. As an example, if we compare the 100 pairs of out-of-control run length values corresponding to **C** and **B1** in case I when $\delta = 1$ and $j_{max} = 50$ (cf. the last column in Table 3), the paired $t$-test for equality of means yields a $p$-value that is less than 0.001. Similar tests conclude that **B2** and **B3** both outperform **C** significantly in this case, and pairwise differences among **B1**, **B2**, and **B3** are not significant in terms of their $ARL_1$ values. More pairwise comparison results are available from the authors upon request.

The general picture that emerges from the above simulations is that, if both $F$ and $G$ are normal, then the conventional CUSUM **C** is a good performer. When the normality does not hold, it can have a too high or too low actual in-control $ARL$ value. The nonparametric procedures **NP1** and **NP2** do not perform well in most cases considered here, due to various reasons, some of which have been discussed in Section 3. In comparison, the proposed bootstrap method does not require prior information about $F$ and $G$. Therefore, it performs reasonably well in all cases considered, as long as $j_{\max}$ is not too small.

**5. An application to aluminum smeltering data.** In this section we consider an example associated with the aluminum smeltering process. The data over 189 time units are on three variables: Silica ($SiO_2$), Ferric Oxide ($Fe_2O_3$), and Magnesium Oxide (MgO), which are denoted as $x_1, x_2$, and $x_3$ below. All these variables are affected by the raw materials, and are relevant to the operation of the smelter. For effective extraction of aluminum, it is desirable that levels of these variables remain stable over time.

Like many other phase II SPC procedures, our procedure assumes that observations at different time points are independent of each other. However, for this dataset, we found that observations at different time points are actually correlated. In the literature, there are several existing discussions regarding SPC procedures in such cases [e.g., Lu and Reynolds (1999), Scariano and Hebert (2003), Zhang (1998)]. A convention is to pre-whiten the observed data by removing the autocorrelation, so that the pre-whitened data may be treated as nearly independent over different time points. In particular, Lu and Reynolds (1999) showed that, for each variable, the original observations have a shift in the mean at a given time point *if and only if* the pre-whitened observations have a shift in the mean at the same time point. When the autocorrelation involved in the data is relatively weak and the potential shift is relatively small, Zhang (1998) suggested an exponentially weighted moving average control chart that is appropriate for original correlated data.

In this example we pre-whiten the observed data by modeling the autocorrelation for each variable with the following $r$th order autoregression model, using the R function ar.yw():

$$(6) \qquad x(i) - \mu = \alpha_1(x(i-1) - \mu) + \cdots + \alpha_r(x(i-r) - \mu) + \varepsilon(i),$$

where $x(i)$ is the measure at the $i$th time point, $\mu$ is its mean, $\alpha_1, \ldots, \alpha_r$ are coefficients, and $\varepsilon(i)$ is a white noise process with zero mean and variance $\sigma_\varepsilon^2$. The default Akaike's Information Criterion (AIC) is used for determining the value of $r$. The results are summarized in Table 6. Residuals from the three fitted autoregression models (denoted as $\varepsilon_1, \varepsilon_2$, and $\varepsilon_3$), corresponding to three original variables $x_1, x_2$, and $x_3$, are then monitored for possible changes in the distributions of $x_1, x_2$, and $x_3$. Figure 1 presents the density curves of the residuals, along with their Normal Q–Q plots. It can be seen that residuals for $x_1$ are right-skewed, those for $x_2$ are a little right-skewed, and those for $x_3$ are a little left-skewed with several small modes besides a major mode around 0. We performed Shapiro–Wilk tests [Shapiro and Wilk (1965)] on these variables to check for normality. The $p$-values are respectively $10^{-16}$, 0.02, and 0.23. This suggests that only residuals for $x_3$ may be approximated with a Normal distribution, in which case we stand to ignore the multiple modes that are present in its density plot. Residuals of the other two variables are significantly non-Normal.

TABLE 6
*Results from the autoregression modeling of the three variables*

| Variable | $\mu$ | $r$ | $\alpha_1, \ldots, \alpha_r$ |
|----------|-------|-----|------------------------------|
| $x_1$ | 0.63 | 3 | 0.07, 0.12, 0.28 |
| $x_2$ | 24.81 | 2 | 0.30, 0.24 |
| $x_3$ | 12.97 | 1 | 0.55 |

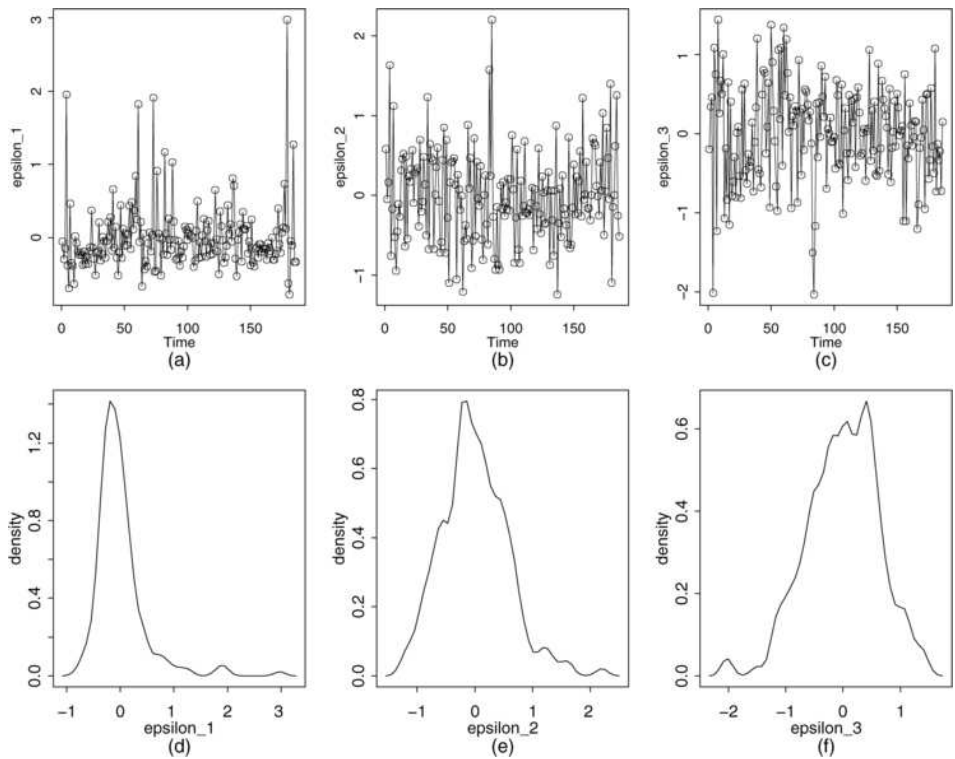

FIG. 1.  *Panels* (a), (b), *and* (c) *present the residuals of the variables* $SiO_2$, $Fe_2O_3$, *and* $MgO$, *respectively, in the aluminum smeltering data, after autocorrelation of the variables were excluded by a autocorrelation model. Panels* (d), (e), *and* (f) *show the corresponding density curves of the residuals.*

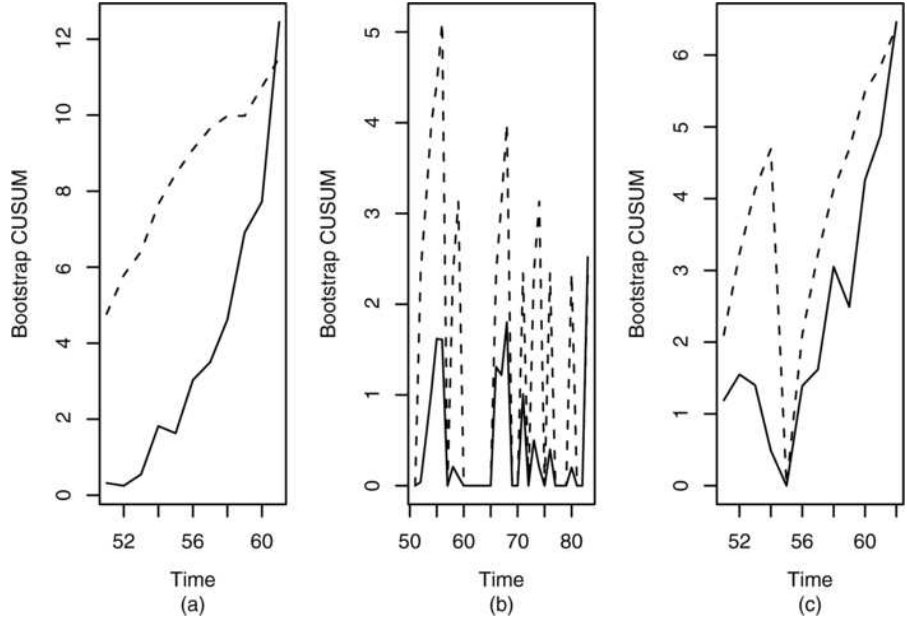

FIG. 2.    *Panels* (a), (b), *and* (c) *present the control charts of the bootstrap CUSUM* **B2** *for* $\varepsilon_1, \varepsilon_2$, *and* $\varepsilon_3$. *The solid line is the CUSUM, and the broken line in the plots denote control limits used at the corresponding time points.*

From the simulation examples presented in Section 4, we notice that the bootstrap CUSUM **B2** performs reasonably well in all cases considered there. To illustrate the use of this procedure in the present application, for each residual variable, we used the first 50 data points as phase I data, and then monitored the remaining observations for possible shifts in $F$. As discussed in Section 2, we first estimated $F$ by $\hat{F}$, from the phase I data, using kernel density estimation. Then the sequence of control limits $\{h_j, j = 1, \ldots, j_{\max}; h^*\}$ were determined by the algorithm described in Section 2, using the bootstrap, where $h^*$ is fixed at 50, $ARL_0 = 200$, and all other procedure parameters are specified as in Section 4. Next, we tried to monitor the 51st–186th observations of each residual variable. The control charts, up to the times when out-of-control was signaled, are shown in Figure 2. From the plots, it can be seen that control limits in the proposed bootstrap procedure at different time points could be different, and they are reset to zero whenever the CUSUM statistic is zero. From the three plots of this figure, it can be seen that we have signals of shifts in $F$ at the 61st, 83rd, and 62nd time points, respectively, for $x_1$, $x_2$, and $x_3$. We also tried bootstrap CUSUMs **B1** and **B3**. Their results are similar. Analysis at subsequent time points show that the CUSUM for silica is increasing almost with a linear trend, while those for Ferric Oxide and Magnesium Oxide are sometimes within the con-trol limits, and sometimes outside. This suggests that there is a discernible pattern change in the silica content of the aluminum ore; with changes in Ferric and Magnesium impurities also being very likely.

Aluminum smelting is an energy intensive, continuous process, and a smelter cannot easily be stopped and restarted. If the production is interrupted for more than four hours, the metal in the pots solidify, often requiring an expensive rebuilding process. See http://www.world-aluminium.org/About+Aluminium/Production/Smelting/index.html for further details. In view of the possible lack of Normality of the data, and considering the enormous cost of a process failure, it is worthwhile to monitor for SPC using the proposed bootstrap based procedure. Since the data are multivariate in nature, another possible approach to monitor this data would be to use a multivariate distribution-free control chart, such as the ones by Qiu (2008) and Qiu and Hawkins (2001, 2003).

**6. Concluding remarks.** In this paper we put forth two proposals: (i) use of a sequence of control limits $\{h_j\}$ which depend on the conditional distributions of $\{C_n | T_n = j\}$, and (ii) obtaining the control limits $\{h_j\}$ by bootstrap. These two proposals result in SPC procedures that do not depend on the in-control distribution $F$ and the out-of-control distribution $G$. Its control limits $\{h_j\}$ are obtained in a data-driven way from a Phase I dataset. Simulation experiments in Section 4 and a real-data example in Section 5 illustrate that it works reasonably well in various cases.

In this paper we suggest choosing $j_{\max}$ from computational considerations, and $\mathbb{E}T_n$ by linking it to $j_{\max}$, as a matter of convenience. The distribution of $(C_n, T_n)$ depends on $(j_{\max}, k)$ in a way that is poorly understood at present. Simulation examples suggest that the best choice of $j_{\max}$ may depend on both $F$ and $G$, however, the convenient choices of $j_{\max}$ used in this paper perform reasonably. More research is required to provide guidelines on selection of $j_{\max}$ and $k$ for bootstrap based SPC.

In this paper we focus on detecting potential shifts in the mean of $F$. Studies are needed on the ability of the proposed method in detecting shifts in variance or other summary statistics of $F$. Also, after obtaining a signal from the proposed procedure, we know that there is a potential shift in $F$; but we do not know whether the shift is in the mean, variance, or any other aspects of $F$. To further investigate this, a possible approach is to apply a SPC procedure for detecting shifts in the mean (e.g., the Shewhart chart), for instance, if we are interested in knowing whether there is a shift in the mean, after obtaining a signal from the proposed method. More research is needed for studying the theoretical and numerical properties of such combined procedures, and for comparison between them and the combined procedures of traditional CUSUMs and Shewhart charts that are commonly used in practice. Extension of the proposed method to multidimensional cases is also a future research topic.

SCHOOL OF STATISTICS
UNIVERSITY OF MINNESOTA
313 FORD HALL
224 CHURCH ST. SE
MINNEAPOLIS, MINNESOTA 55455
USA
E-MAIL: chatterjee@stat.umn.edu
         qiu@stat.umn.edu